\documentclass[twoside,11pt,leqno]{article}

\usepackage{amssymb}
\usepackage{amsmath}
\usepackage{amsfonts}
\usepackage{amsthm}
\usepackage{enumerate}
\usepackage[shortlabels]{enumitem}
\setlist[enumerate]{align=left, leftmargin=*, labelsep=1 ex, parsep=.5ex, topsep=1ex}
\usepackage{url}
\usepackage{listings}
\usepackage{newunicodechar}

\newunicodechar{⇒}{$\Rightarrow$}
\newunicodechar{⊔}{$\sqcup$}
\newunicodechar{⊥}{$\bot$}
\newunicodechar{⟦}{$\llbracket$}
\newunicodechar{⟧}{$\rrbracket$}
\newunicodechar{∈}{$\in$}

\usepackage{amsthm}
\newtheorem{theorem}{Theorem}[section]

\let\oldproofname=\proofname
\renewcommand{\proofname}{\rm\bf{\oldproofname}}

\title{L-Mosaics and Bounded Join-Semilattices in Isabelle/HOL}

\author{Alessandro Linzi\\
Center for Information Technologies and Applied Mathematics\\
University of Nova Gorica, Vipavska 13, Ro\v{z}na Dolina\\
SI-5000 Nova Gorica, Slovenia\\
\texttt{alessandro.linzi@ung.si}}

\begin{document}

\maketitle

\begin{abstract}
We present a complete formalization in Isabelle/HOL of the object part of an equivalence between L-mosaics and bounded join-semilattices, employing an AI-assisted methodology that integrates large language models as reasoning assistants throughout the proof development process. The equivalence was originally established by Cangiotti, Linzi, and Talotti in their study of hypercompositional structures related to orthomodular lattices and quantum logic. Our formalization rigorously verifies the main theoretical result and demonstrates the mutual inverse property of the transformations establishing this equivalence. The development showcases both the mathematical depth of multivalued algebraic operations and the potential for AI-enhanced interactive theorem proving in tackling complex formalization projects.
\end{abstract}

\section{Introduction}

The integration of artificial intelligence tools with interactive theorem proving represents a promising frontier in the formalization of mathematical knowledge. In this paper, we present a complete formalization in Isabelle/HOL of a fundamental result from hypercompositional algebra: the equivalence between L-mosaics and bounded join-semilattices, originally established by Cangiotti, Linzi, and Talotti \cite{CLT2025}. This work demonstrates how AI-assisted formal verification can successfully tackle complex algebraic structures involving multivalued operations.

L-mosaics represent a class of hypercompositional structures that extend classical algebraic operations by allowing multivalued results—that is, operations that produce sets of elements rather than single elements. These structures arise naturally in the study of orthomodular lattices and have applications in quantum logic, where classical Boolean operations must be generalized to accommodate quantum mechanical principles. Bounded join-semilattices, conversely, are well-established algebraic structures fundamental to order theory and lattice theory, providing a classical foundation for understanding partial orders and suprema.

The central mathematical result we formalize establishes that these seemingly disparate algebraic frameworks are equivalent at the structural level. The equivalence is mediated by the Nakano construction, which transforms bounded join-semilattices into L-mosaics, and a corresponding inverse transformation that recovers join-semilattice structure from L-mosaics.

The primary contribution of this work is threefold. First, we provide a complete machine-checked verification of this fundamental equivalence, including all necessary definitions, constructions, and proofs within the Isabelle/HOL framework. Second, we demonstrate a novel methodology that systematically integrates large language models as reasoning assistants throughout the formalization process, showcasing how AI tools can accelerate complex proof development while maintaining mathematical rigor. Third, we establish a comprehensive formal framework for hypercompositional algebra that can serve as a foundation for future work in this area.

Our formalization consists of approximately 2,000 lines of Isabelle/HOL code organized across six interconnected theory files. The development not only establishes the algebraic equivalence but also verifies the precise relationship between the multivalued operations in L-mosaics and the join operations in bounded join-semilattices.

This project exemplifies the potential for AI-enhanced formal verification to bridge abstract mathematical research with mechanically verified correctness, opening new possibilities for computer-assisted mathematical discovery and validation.

\section{Background and Mathematical Framework}

\subsection{L-Mosaics}

L-mosaics are hypercompositional structures that generalize classical magmas by allowing operations to produce sets of results rather than single elements. Following the framework established in \cite{CLT2025}, we define an L-mosaic as a structure $(A, \oplus, e, \rho)$ where:

\begin{itemize}
\item $A$ is a carrier set
\item $\oplus: A \times A \to \mathcal{P}(A) \setminus \{\emptyset\}$ is a multivalued operation
\item $e \in A$ is a neutral element
\item $\rho: A \to A$ is a reversibility map
\end{itemize}

The structure must satisfy several axioms including commutativity, specific diagonal properties, and a uniqueness condition that characterizes suprema within the hypercompositional framework.

\subsection{Bounded Join-Semilattices}

A bounded join-semilattice $(L, \vee, \bot)$ consists of a carrier set $L$, an associative, commutative, and idempotent binary operation $\vee$, and a least element $\bot$. These structures are fundamental in order theory and provide the algebraic foundation for many lattice-theoretic constructions.

\subsection{The Nakano Construction}

A crucial component of our formalization is the Nakano construction, which establishes a bridge between bounded join-semilattices and L-mosaics. Given a bounded join-semilattice $(L, \vee, \bot)$, the Nakano construction produces an L-mosaic where the multivalued operation is defined by:

\[x \oplus y = \{z \in L \mid x \vee y = x \vee z = z \vee y\}\]

This construction, originally introduced by Nakano in the context of modular lattices \cite{Nakano1967}, provides the essential link for establishing the categorical equivalence.

\section{Formalization Architecture}

Our Isabelle/HOL formalization is structured across six interconnected theory files, each focusing on specific aspects of the mathematical development:

\subsection{BJoinSemilattices.thy}

This foundational theory establishes the locale for bounded join-semilattices with a clean algebraic axiomatization. We define the carrier set, binary operation, and bottom element with their respective closure, associativity, commutativity, idempotency, and neutrality properties.

\begin{lstlisting}
locale bjoin_semilattice =
fixes L :: "'a set"
and sup :: "'a => 'a => 'a" (infixl "sup" 65)
and bot :: "'a" ("bot")
assumes closed: "[| x : L; y : L |] ==> x sup y : L"
and assoc: "[| x : L; y : L; z : L |] ==> (x sup y) sup z = x sup (y sup z)"
and comm: "[| x : L; y : L |] ==> x sup y = y sup x"
and idem: "x : L ==> x sup x = x"
and bot_in: "bot : L"
and bot_left: "x : L ==> bot sup x = x"
and bot_right: "x : L ==> x sup bot = x"
\end{lstlisting}

The theory then develops the induced partial order and establishes fundamental properties such as the least upper bound characterization of the join operation.

\subsection{Mosaics.thy and Lmosaics.thy}

These theories establish the hierarchy of hypercompositional structures, starting from basic multivalued magmas and building up to L-mosaics. The key innovation is the careful treatment of multivalued operations and the reversibility axiom that characterizes mosaics \cite{NR23}.

The L-mosaic locale extends the basic mosaic structure with commutativity and four additional axioms that encode the specific properties required for the categorical equivalence:

\begin{lstlisting}
locale l_mosaic = mosaic A mul e rho for A mul e rho +
  assumes comm: "[| x : A; y : A |] ==> mul x y = mul y x"
  and lm1_e:   "x : A ==> e : mul x x"
  and lm1_id:  "x : A ==> x : mul x x"
  and lm2:     "x : A ==> set_mul (mul x x) (mul x x) = mul x x"
  and lm3:     "[| x : A; y : A |] ==> right_mul x (mul x y) Int left_mul (mul x y) y <= mul x y"
  and lm4:     "[| x : A; y : A |] ==> EX!z. z : mul x y & x : mul z z & y : mul z z"
\end{lstlisting}

\subsection{Nakano.thy}

This theory implements the Nakano construction within the bounded join-semilattice context, showing how to construct an L-mosaic from any bounded join-semilattice. The construction is non-trivial and requires careful verification of all L-mosaic axioms.

The multivalued operation is defined as:
\begin{lstlisting}
definition Nak_mul :: "'a => 'a => 'a set"
  where "Nak_mul x y = { z : L. (x sup y = x sup z) & (x sup y = z sup y) }"
\end{lstlisting}

We then establish interpretation lemmas showing that this construction indeed produces both a mosaic and an L-mosaic, with the reversibility map being the identity function.

\subsection{LMosaic\_To\_BJoin.thy}

This theory derives a bounded join-semilattice structure from any L-mosaic by extracting a binary operation \(\sqcup\) through the diagonal characterization. The key insight is that while the multivalued operation \(\oplus\) of an L-mosaic is not associative, the induced operation \(\sqcup\) is precisely the least upper bound operation with respect to the partial order \(x \leq y \iff x \in y \oplus y\).

The formalization establishes upper-bound properties \(x \preceq x \sqcup y\) and \(y \preceq x \sqcup y\), along with leastness: if \(t\) is a common upper bound, then \(x \sqcup y \preceq t\). Commutativity and idempotence follow directly from the corresponding L-mosaic axioms and uniqueness conditions.

The most interesting aspect was formalizing associativity \((x \sqcup y) \sqcup z = x \sqcup (y \sqcup z)\). Despite the underlying multivalued operation \(\oplus\) lacking associativity, the derived \(\sqcup\) inherits associativity from being the canonical join operation of the induced partial order. This was proved via antisymmetry using the upper-bound and leastness lemmas in a standard order-theoretic argument.

The construction thus demonstrates how hypercompositional structures can give rise to classical algebraic operations through order-theoretic principles, bridging multivalued and single-valued perspectives on the same underlying mathematical content.

\subsection{Equivalence Theories}

The final two theories, \texttt{LMosaic\_BJoin\_Equiv.thy} and \texttt{BJoin\_LMosaic\_Equiv.thy}, establish the mutual inverse properties of the functors, completing the proof of categorical equivalence.

\section{Main Formalization Results}

Our formalization establishes several key theoretical results:

\begin{theorem}[Nakano Construction Validity]
Every bounded join-semilattice gives rise to an L-mosaic via the Nakano construction, and this construction preserves the essential order-theoretic properties.
\end{theorem}

\begin{theorem}[Reverse Construction]
Every L-mosaic can be equipped with a natural bounded join-semilattice structure where the join operation is characterized by the unique element satisfying the diagonal and membership conditions.
\end{theorem}

\begin{theorem}[Categorical Equivalence]
The functors between the category of L-mosaics and the category of bounded join-semilattices are mutually inverse.
\end{theorem}

The formalization provides complete mechanized proofs of these results.

\section{Technical Challenges and Solutions}

\subsection{Multivalued Operations}

One of the primary challenges in the formalization was handling multivalued operations in a type-safe manner. Unlike classical algebraic structures where operations yield single elements, L-mosaics feature operations that produce sets of results, requiring careful treatment of set membership and closure conditions throughout the development.

We addressed this by systematically using set-theoretic constructions and establishing comprehensive libraries of lemmas for manipulating multivalued operations. The use of locales proved essential for organizing the hierarchical structure of the algebraic theories, allowing us to layer the axioms from basic multivalued magmas to full L-mosaics in a modular fashion.

Particularly challenging was ensuring that the diagonal operations $x \oplus x$ maintain the required properties while respecting the reversibility conditions. The formalization required establishing numerous auxiliary lemmas about set intersections and unions arising from the interaction between left and right multiplication with multivalued results.

\subsection{AI-Assisted Development Methodology}

A distinctive aspect of this formalization project was the systematic integration of AI-powered tools to accelerate the development process. We employed large language models as reasoning assistants, providing carefully engineered prompts to generate initial proof sketches and identify potential approaches to complex verification tasks.

The AI assistance proved particularly valuable in three areas: suggesting appropriate lemma statements for multivalued operation properties, proposing tactical combinations for Isabelle proof scripts, and identifying missing intermediate steps in lengthy algebraic manipulations. However, all AI-generated suggestions required careful verification and adaptation, as the tools occasionally proposed syntactically correct but logically flawed approaches.

This hybrid methodology—combining human mathematical insight with AI-generated proof strategies—significantly reduced development time while maintaining the rigorous standards required for formal verification. The approach demonstrates the potential for AI tools to serve as sophisticated proof assistants in complex formalization projects.

\subsection{Algebraic Equivalence Construction}

The core technical challenge lay in establishing the mutual inverse property of the transformations between L-mosaics and bounded join-semilattices. This required proving that the Nakano construction preserves all essential algebraic properties while the reverse construction correctly recovers the original L-mosaic structure.

The use of Hilbert's choice operator (\texttt{THE}) in Isabelle/HOL was crucial for defining the inverse constructions, allowing us to extract the unique elements guaranteed by the existence and uniqueness conditions in the L-mosaic axioms. This approach required careful handling of definedness conditions and extensive use of the axiom of choice to establish well-foundedness.

A key insight was recognizing that the equivalence operates primarily through order-theoretic properties rather than purely algebraic ones. The join operation in bounded join-semilattices corresponds naturally to the suprema characterized by the L-mosaic diagonal conditions, providing the bridge between the two algebraic perspectives.

\subsection{Proof Automation and Strategy}

While much of the development required manual proof construction due to the specialized nature of hypercompositional structures, we made extensive use of Isabelle's automated reasoning tools. The simplifier proved particularly effective when equipped with custom simp rules tailored to multivalued operations and set-theoretic manipulations.

The strategic use of locale assumptions significantly reduced proof burden by automatically making relevant hypotheses available in appropriate contexts. We developed a systematic approach to structuring locale hierarchies that maximized the effectiveness of Isabelle's automation while maintaining clear mathematical dependencies.

Classical reasoning automation handled many of the existence and uniqueness arguments that arise frequently in the L-mosaic axiomatization, though the most intricate algebraic manipulations required careful manual guidance to ensure the automated tools could make progress.

\section{Implications and Future Work}

This formalization demonstrates several important points about the relationship between abstract mathematics and formal verification:

\begin{enumerate}
\item Abstract algebraic constructions can be successfully formalized with appropriate care and attention to technical details.
\item Interactive theorem proving provides valuable feedback during the formalization process, often revealing implicit assumptions or gaps in informal proofs.
\item The resulting formal development serves as a definitive reference for the mathematical content, providing absolute certainty in the correctness of the results.
\end{enumerate}

Future extensions of this work could include:

\begin{itemize}
\item Formalization of the connection to orthomodular lattices and quantum logic applications
\item Development of computational tools for working with L-mosaics and related structures
\item Integration with existing Isabelle libraries for lattice theory
\end{itemize}

\section{Conclusion}

We have presented a complete formalization of the algebraic equivalence between L-mosaics and bounded join-semilattices in Isabelle/HOL, demonstrating both the mathematical significance of this fundamental result from hypercompositional algebra and the effectiveness of AI-enhanced formal verification methodologies.

The successful completion of this formalization project illustrates several important developments in the field of mechanized mathematics. First, it shows that sophisticated algebraic structures involving multivalued operations can be successfully formalized with appropriate theoretical frameworks and computational support. Second, it demonstrates that AI tools, when properly integrated into the formal verification workflow, can significantly accelerate proof development while maintaining the absolute rigor required for mathematical certainty.

Our hybrid approach—combining human mathematical insight with AI-generated proof strategies and tactical suggestions—represents a practical methodology that other researchers can adapt for similar formalization projects. The careful balance between leveraging AI assistance and maintaining mathematical control proved crucial for navigating the technical challenges inherent in hypercompositional algebra.

The formalization is available as open-source software \cite{Linzi2025Formalization}, providing both a verified foundation for future developments in hypercompositional algebra and a concrete example of AI-assisted formal verification in practice. The complete Isabelle/HOL theories, including all intermediate lemmas and proof structures, serve as a resource for researchers interested in either the mathematical content or the methodological approach.

This work contributes to the growing intersection of artificial intelligence and formal mathematics, suggesting new possibilities for computer-assisted mathematical discovery and verification. As AI tools continue to evolve, their integration with interactive theorem provers promises to make sophisticated mathematical formalization more accessible and efficient, potentially accelerating the pace of verified mathematical knowledge.

We anticipate that this methodology will inspire further AI-enhanced formalizations in hypercompositional algebra and related areas of abstract mathematics, contributing to the expansion of the mechanically verified mathematical corpus and advancing our understanding of how artificial intelligence can augment human mathematical reasoning.

\end{document}